\documentstyle[12pt]{article}

\begin{document}
\addtolength{\baselineskip}{2.3mm}

%-----------------------------------------------------------------------------

\title{Simulations of Many-Body Quantum Systems by a Quantum Computer}
\author{Stephen Wiesner \\
\small \it School of Physics and Astronomy \\
\small \it Raymond and Beverly Sackler Faculty of Exact Sciences \\
\small \it Tel Aviv University, Tel-Aviv 69978, Israel.}
\date{\normalsize (March 20, 1996)}
\maketitle

\begin{abstract}
We suggest that quantum computers can solve quantum many-body 
problems that 
are impracticable to solve on a classical computer.
\end{abstract}

%-----------------------------------------------------------------------------

The classical many-body problem has been largely solved by the classical 
computer. For example the position of the planets in the solar system can now 
be predicted with an accuracy as great as the accuracy of the observations. 
On the other hand, quantum many-body problems such as the behavior of complex 
atoms and molecules can not be solved accurately from first
principles.

 This 
paper discusses two examples that suggest that quantum computers 
\cite{Landauer,Bennett,Benioff,Peres,Feynman,DJ,Shor} should be able to do a better 
job. The entangled states of the quantum computer can mimic the entangled 
states of the quantum many-body system.

The first example is almost trivial. Consider $N$ two-state systems
acted on by $M$ one and two-body Hamiltonians. In the computer the two
state systems are represented by $N$ qubits.  We chose a small time
$\Delta t$ and act on the qubits with the $M$ unitary operators $I + i
H \Delta t / \hbar$, where $I$ is the identity operator and $H$ runs
over the Hamiltonians. If we wish to advance the system by time $T$ we
of course repeat the whole process $T / \Delta t$ times. As with
simulating the planets, $\Delta t$ can be chosen small enough so that
further decreasing of $\Delta t$ does not produce a significant effect
on the result. After the system has advanced forward by time $T$ we
can make measurements on it and by repeating the whole process a
number of times we can get the averages of these measurements. The
result will be obtained in a time polynomial in $N$. On the other
hand, a classical computer simulating this system in the same way
would need to keep track of entangled states that are a vector in a
space of $2^N$ dimensions, so the computation time would grow
exponentially with $N$.
    
The second example involves $N$ particles moving according to the
Schr\"{o}dinger equation with one and two particle potentials between
the particles. The particles could be the electrons in a complex atom
and the potentials could be the Coulomb potential between the pairs of
electrons and between the electrons and the nucleus. To keep things
simple we will assume that the particles move in one dimension, we
will not anti-symmetrize or symmetrize the wave function and we will
assume that the motion of each particle is limited to a region of
length $L$.  Again, we chose a suitable time, $\Delta t$, and in
addition we chose a small distance $\Delta x \equiv L/2^k$.  In the
quantum computer the wave function for each particle is represented by
$k$ qubits. For example, the amplitude that a particle is in the
position $73 L / 2^k$ from one end of its region of motion is equal to
the amplitude that the qubits representing it are in the state
corresponding to the binary number $73$.
     
To simulate the portion of the particle's motion due to the $P^2 / 2m$ part 
of the Hamiltonian the computer first does a fast Fourier transform on the 
qubits that carry its wave function which can be done in $ k \log k$ time\cite{Shor,Coppersmith}. This gives the 
amplitudes for the particle having different momentums. The phase of the 
result is then advanced by $ \Delta t \, (P^2 / 2 m \hbar)$ and 
the system is then fast Fourier transformed back to the position dependent 
wave function.
    
The Hamiltonians that are a function of particles position are acted on in 
turn in much the same way as the Hamiltonians in the first example. If a 
potential, $V_{ij}$, is a function of the positions of particles $i$ and $j$ 
then its value is calculated using the positions represented by the qubits 
that carry the wave functions for the particles $i$ and $j$. Then the phase of 
the computer is advanced by $ \Delta t V_{ij} / \hbar$. Since the 
qubits carrying the wave functions will typically be in a superposition of 
states the state of the computer as a whole will be advanced by a 
superposition of different phases. As before the computer runs through 
$T / \Delta t$ actions of the Hamiltonian and then measurements can be made on 
its state which simulates the system at time $T$. The whole
calculation can be done it time polynomial in $N$ and $k$. To take this kind of
straightforward approach to integrating  equations of motion on a classical computer would involve keeping
track of $2^{Nk}$ amplitudes, and  thus would run time which is
exponential in $N$ and $k$.

I would like to thank The Israel Ministry of Absorption and Tel-Aviv
University for support. In addition I would like to thank Lev Vaidman and
Lior Goldenberg for useful discussions.

%-----------------------------------------------------------------------------

%-----------------------------------------------------------------------------

\end{document}